# Klein-Gordon Equation with Casimir Potential for Attoseconds Laser Pulse Interaction with Matter

Mirosław Kozłowski[*] Janina Marciak-Kozłowska

Institute of Electron Technology, Al. Lotników 32/46, Warsaw Poland

---

[*] Corresponding author, e-mail: MiroslawKozlowski@aster.pl

**Abstract**

In this paper the thermal Klein-Gordon equation (K-GE) is solved for the interaction of attosecond laser pulses with medium in which Casimir force operates. It is shown that for nanoscale structures, NEMS and MEMS, the attosecond laser pulses can be used as the tool for the investigation of the role played by Casimir force on the nanoscale.

**Key words**: Casimir force; NEMS, MEMS, Attosecond laser pulses.

## 1   Introduction

The contemporary nanoelectronic develops the NEMS and MEMS structures in which the distance between the parts is of the order of nanometers. As was shown in our monograph: *From quarks to bulk matter* the transport phenomena on the nanoscale dependent on the second derivative in time. It is in contrast to the macroscale heat transport where the Fourier law (only the first derivative in time). The second derivative in time term describes the memory of the thermally excited medium. Considering the contemporary discussion of the role played by Casimir force in the NEMS and MEMS in this paper we describe the heat signaling in the simple nanostructure the parallel plates heated by attosecond laser pulses. It will be shown that temperature field between plates depends on the distance of the plates (in nanoscale). As the result the attosecond laser pulse can be used as the tool for the investigation of Casimir effect on the performance of the NEMS and MEMS.

## 2   Repulsive quantum vacuum forces

Vacuum energy is a consequence of the quantum nature of the electromagnetic field, which is composed of photons. A photon of frequency ω has energy $\hbar\omega$, where $\hbar$ is Planck constant. The quantum vacuum can be interpreted as the lowest energy state (or ground state) of the electromagnetic (EM) field that occurs when all charges and currents have been removed and the temperature has been reduced to absolute zero. In this state no ordinary photons are present. Nevertheless, because the electromagnetic field is a quantum system the energy of the ground state of the EM is not zero. Although the average value of the electric field $\langle E \rangle$ vanishes in ground state, the Root Mean Square of the field $\langle E^2 \rangle$ is not zero. Similarly the $\langle B^2 \rangle$ is not zero. Therefore the electromagnetic field energy, $\langle E^2 \rangle + \langle B^2 \rangle$ is not equal zero. A detailed theoretical calculation tells that EM energy in each mode of oscillation with frequency ω is $0.5\hbar\omega$, which equals one half of the amount of energy that would be present if a single "real" photon of that mode were present. Adding up $0.5\hbar\omega$ for all possible modes of the electromagnetic field gives a very large number for the vacuum energy $E_0$ in the quantum vacuum

$$E_0 = \sum_i \frac{1}{2}\hbar\omega_i . \qquad (1)$$

The resulting vacuum energy $E_0$ is *infinity* unless a high frequency cut off is applied.

Inserting surfaces into the vacuum causes the modes of the EM to change. This change in the modes that are present occurs since the EM must meet the appropriate boundary conditions at each surface. Surface alter the modes of oscillation and therefore the surfaces alter the energy density corresponding to the lowest state of the EM field. In actual practice the change in $E_0$ is defined as follows

$$\Delta E_0 = E_0 - E_S, \qquad (2)$$

where $E_0$ is the energy in empty space and $E_S$ is the energy in space with the surfaces, i.e.

$$\Delta E_0 = \frac{1}{2}\sum_n^{\text{empty space}} \hbar\omega_n - \frac{1}{2}\sum_i^{\text{surface present}} \hbar\omega_i . \qquad (3)$$

As an example let us consider a hollow conducting rectangular cavity with sides $a_1, a_2, a_3$. In that case for uncharged parallel plates with area $A$ the attractive force between the plates is, [1]:

$$F_{att} = -\frac{\pi^2 \hbar c}{240 d^4} A, \qquad (4)$$

where $d$ is the distance between plates. The force $F_{att}$ is called the parallel plate Casimir force, which was measured in three different experiments [2, 3, 4].

Recent calculations show that for conductive rectangular cavities the vacuum forces on a given face can be repulsive (positive), attractive (negative) or zero depending on the ratio of the sides [5].

In paper [6] the first measurement of repulsive Casimir force was performed. For the distance (separation) $d \sim 0.1\,\mu\text{m}$ the repulsive force is of the order of $0.5\,\mu\text{N}$ (micronewton) – for cavity geometry. In March 2001, scientist at Lucent Technology used attractive parallel plate Casimir force to actuate a MEMS torsion device [4]. Other MEMS (MicroElectroMechanical Systems) have been also proposed [7].

### 3     Klein-Gordon thermal equation with Casimir force

Standard Klein-Gordon equation reads:

$$\frac{1}{c^2}\frac{\partial^2 \Psi}{\partial t^2} - \frac{\partial^2 \Psi}{\partial x^2} + \frac{m^2 c^2}{\hbar^2}\Psi = 0. \qquad (5)$$

In equation (1) $\Psi$ is the relativistic wave function for particle with mass $m$, $c$ is the light velocity and $\hbar$ is Planck constant. For massless particles $m = 0$ and Eq. (1) is the Maxwell equation for photons. As was shown by Pauli and Weisskopf since relativistic quantum mechanical equation had to allow for creation and anihilation of particles, the Klein-Gordon describes spin – 0 bosons.

In the monograph by two of us (Janina Marciak-Kozłowska and Mirosław Kozłowski) [8] the generalized Klein-Gordon thermal equation was developed

$$\frac{1}{v^2}\frac{\partial^2 T}{\partial t^2} - \nabla^2 T + \frac{m}{\hbar}\frac{\partial T}{\partial t} + \frac{2Vm}{\hbar^2} = 0. \qquad (6)$$

In Eq. (6) $T$ denotes temperature of the medium and $v$ is the velocity of the temperature signal in the medium. When we extract the highly oscillating part of the temperature field,

$$T = e^{-\frac{t\omega}{2}} u(x,t), \qquad (7)$$

where $\omega = \tau^{-1}$, and $\tau$ is the relaxation time, we obtain from Eq. (3) (1D case)

$$\frac{1}{v^2}\frac{\partial^2 u}{\partial t^2} - \frac{\partial^2 u}{\partial x^2} + qu(x,t) = 0, \qquad (8)$$

where

$$q = \frac{2Vm}{\hbar^2} - \left(\frac{mv}{2\hbar}\right)^2. \qquad (9)$$

When $q > 0$ equation (4) is of the form of the Klein-Gordon equation in the potential field $V(x,t)$. For $q < 0$ Eq. (8) is the modified Klein-Gordon equation. The discussion of the physical properties of the solution of equation (4) can be find in [8].

In the paper we will study the heat signaling in the medium excited by ultra-short laser pulses, $\Delta t < \tau$. In that case the solution of Eq. (1) can be approximated as

$$T(x,t) \cong u(x,t) \qquad \text{for} \qquad \Delta t << \tau. \qquad (10)$$

Considering the existence of the attosecond laser with $\Delta t \approx 1\,\text{as} = 10^{-18}\,\text{s}$, Eq. (8) describes the heat signaling for thermal energy transport induced by ultra-short laser pulses. In the subsequent we will consider the heat transport when $V$ is the Casimir potential. As was shown in paper [6] the Casimir force, formula (4), can be repulsive sign $(V) = +1$ and attractive sign $(V) = -1$. For attractive Casimir force, $V < 0, q < 0$ (formula (5)) and equation (4) is the modified K-G equation. For repulsive Casimir force $V > 0$ and $q$ can be positive or negative.

As was shown by J. Maclay [6] for different shapes of cavities the vacuum Casimir force can be changes the sign. In the subsequent we consider the propagation of thermal wave in the geometry described in Fig. 1. In Fig. 2(a) the value of the parameter $q$ as the function of $d$ – distance between the plates and $\frac{v}{c}$ the ratio of the $v$ – thermal wave velocity and $c$ – the light velocity is presented. Considering that $v \approx 10^{-2} c$, from Fig. 2(b) we conclude that $q$ change the sign for $d = 0.759$ nm. For Cauchy initial condition:

$$u(x,0) = 0, \qquad \frac{\partial u(x,0)}{\partial t} = f(x)$$

the solution of Eq. (8) has the form

$$u(x,t) = \frac{1}{2v} \int_{x-vt}^{x+vt} f(x) J_0 \sqrt{q(v^2 t^2 - (x-\zeta)^2)} d\zeta \qquad \text{for} \qquad q > 0 \qquad (11)$$

and

$$u(x,t) = \frac{1}{2v} \int_{x-vt}^{x+vt} f(x) I_0 \sqrt{-q(v^2 t^2 - (x-\zeta)^2)} d\zeta \qquad \text{for} \qquad q < 0. \qquad (12)$$

In the subsequent we choose as the initial distribution temperature, $f(x)$, viz.

$$f(x) = \text{Exp}[-x^2].$$

In Fig. 3(a, b) we present the solution of equation (8) for $q = 0$. In Fig. 3(a) $u(x,t) = T(x,t)$ is calculated as the solution of the modified K-G. Both solution are the same shape due to the fact that $q = 0$. In Fig. 4(a, b) the solution of K-GE and MK-G equation i.e. for $q > 0$ and $q < 0$ are presented for different value of $q$.

**Conclusion**

In this paper the solution of K-G and MK-G for ultra-short laser pulses are presented. It is shown that for attosecond laser pulses the passage of thermal wave through cavity where Casimir potential operates strongly depends on distance between parallel plates.

**Figure captions**

Fig. 1. The thermal wave and parallel plates.

Fig. 2(a). The *q* parameter as the function of *d* and *v/c*, (b) The *q* parameter as the function of *d*, for $v = 10^{-2} c$.

Fig. 3(a) The solution of equation (8) for *d* = 0.759 nm, *q* = 0. (b) The solution of equation MK-GE (8) for *d* = 0.759 nm, *q* = 0.

Fig. 4(a) The solution of equation K-GE (8) for *d* = 0.720, *q* > 0. (b) The solution of equation MK-GE (8) for *d* = 0.760, *q* < 0.

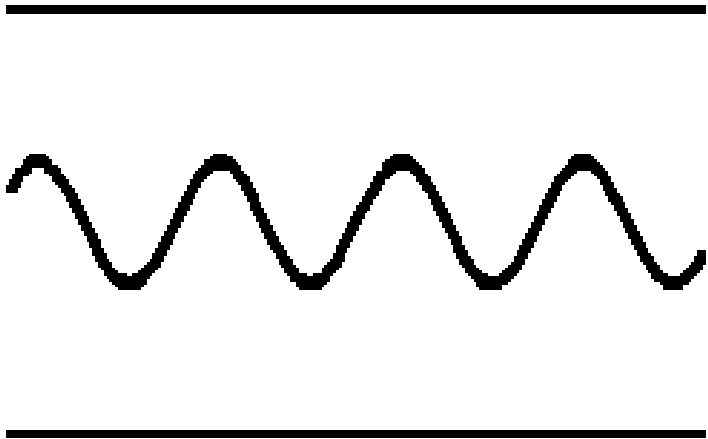

Fig. 1. The thermal wave and parallel plates.

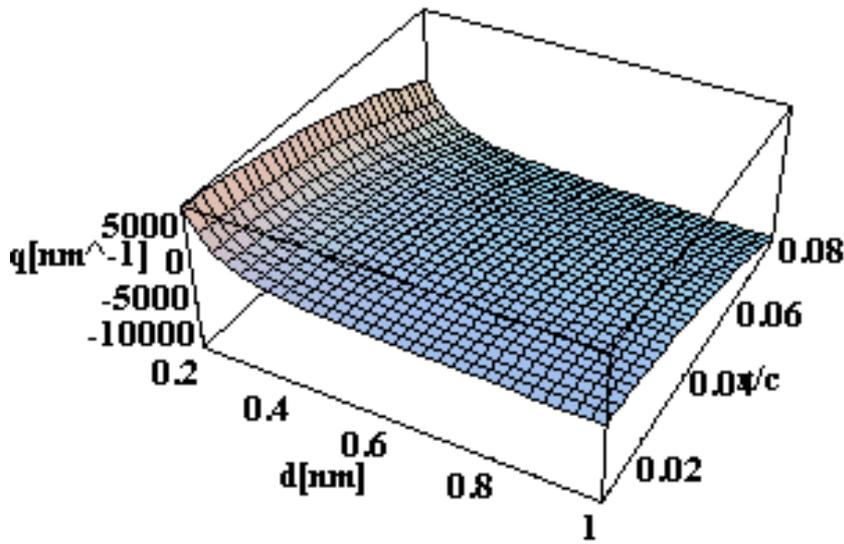

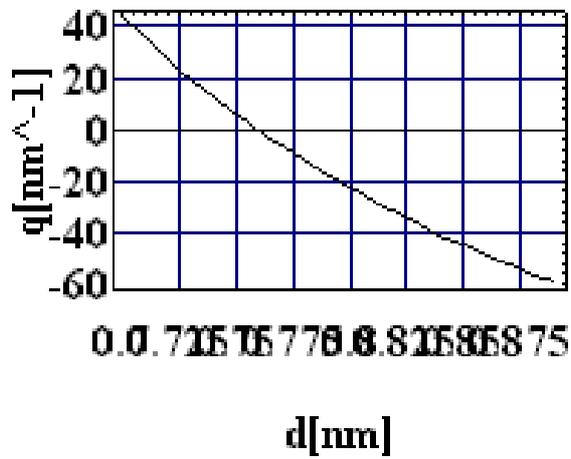

Fig. 2(a). The *q* parameter as the function of *d* and *v/c*, (b) The *q* parameter as the function of *d*, for $v = 10^{-2} c$.

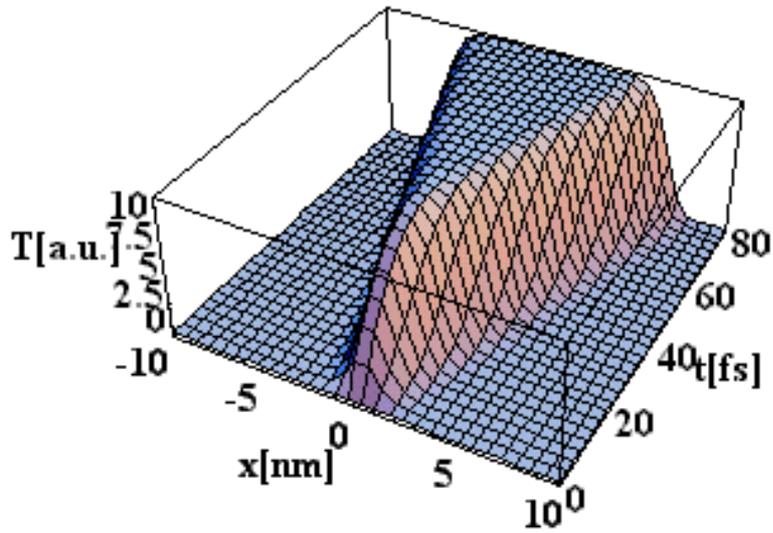

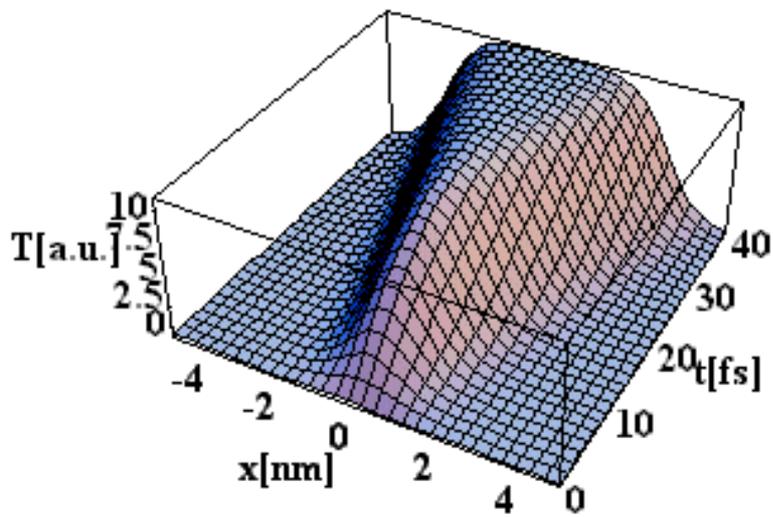

Fig. 3(a) The solution of equation (8) for $d = 0.759$ nm, $q = 0$. (b) The solution of equation MK-GE (8) for $d = 0.759$ nm, $q = 0$.

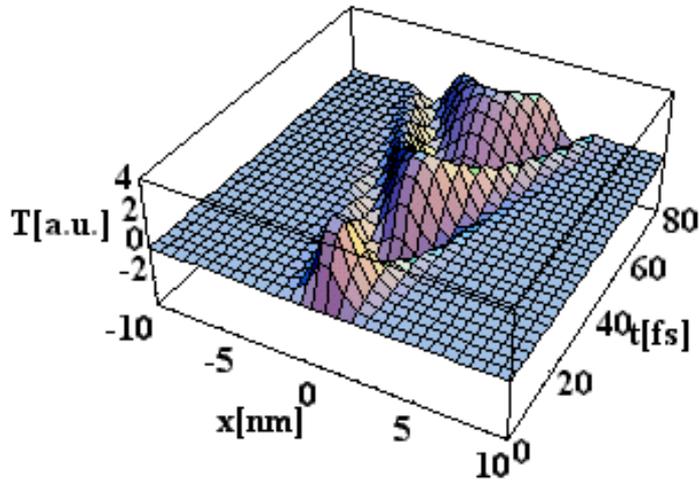

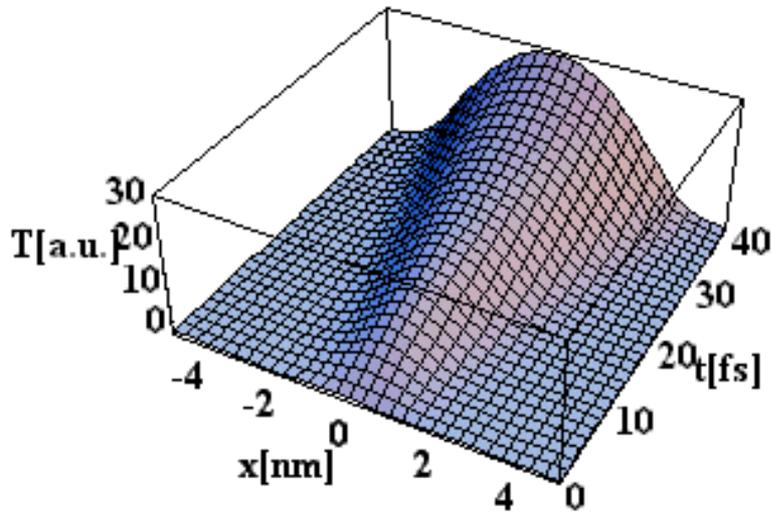

Fig. 4(a) The solution of equation K-GE (8) for $d = 0.720$, $q > 0$. (b) The solution of equation MK-GE (8) for $d = 0.760$, $q < 0$.